\documentclass{Interspeech}
\usepackage{tipa}
\usepackage{array}

\newcolumntype{P}[1]{>{\centering\arraybackslash}p{#1}}
\usepackage{multirow}

\interspeechcameraready

\title{On the Relationship between Accent Strength and Articulatory Features}

\author[affiliation={1}]{Kevin}{Huang}
\author[affiliation={1,2}]{Sean}{Foley}
\author[affiliation={1}]{Jihwan}{Lee}
\author[affiliation={1}]{Yoonjeong}{Lee}
\author[affiliation={2}]{Dani}{Byrd}
\author[affiliation={1}]{Shrikanth}{Narayanan}

\affiliation{Signal Analysis and Interpretation Laboratory}{University of Southern California}{USA}
\affiliation{Department of Linguistics}{University of Southern California}{USA}
\email{kevinyhu@usc.edu}

\usepackage{comment}

\begin{document}

\maketitle

\begin{abstract}
    This paper explores the relationship between accent strength and articulatory features inferred from acoustic speech. To quantify accent strength, we compare phonetic transcriptions with transcriptions based on dictionary-based references, computing phoneme-level difference as a measure of accent strength. The proposed framework leverages recent self-supervised learning articulatory inversion techniques to estimate articulatory features. Analyzing a corpus of read speech from American and British English speakers, this study examines correlations between derived articulatory parameters and accent strength proxies, associating systematic articulatory differences with indexed accent strength. Results indicate that tongue positioning patterns distinguish the two dialects, with notable differences inter-dialects in rhotic and low back vowels. These findings contribute to automated accent analysis and articulatory modeling for speech processing applications.
\end{abstract}

\section{Introduction}

Speech accents involve pronunciation and prosodic differences relative to other dialects of that language. (Here we limit our consideration of accent to dialectal variation driven by geographically associated forms of a one language) 
The degree and nature of these differences among dialects are myriad, influenced by the speakers' lived experiences, their socioeconomic and cultural groups, and their exposure to other dialects. While research on the acoustics of various dialects is rich in its historical and scholarly depth, acoustics are causally shaped by the actions of vocal tract articulators including the tongue, lips, and jaw. Individual articulatory variations and differences in accent strength (or typicality) exist among speakers within the same `nominal' accent group—for example, not all Americans sound the same and likewise not all British sound the same, despite the well known and salient gross dialect differences between the two groups. An examination of the causal articulatory underpinnings of accent may provide insight into the underlying parameters of such variation. This study examines whether and how articulatory features correlate with perceived accent strength. 

British and American English differ notably in rhoticity (e.g., \textit{car} pronounced as /\textipa{kA\textrhookrevepsilon}/ in American English vs. /\textipa{kA:}/ in British English). American English rhotic vowels typically involve a retracted tongue position, achieved through retroflexion (tongue tip raises and curled back) or bunching (tongue root retracted and/or raised)~\cite{Alwan1997Towardarticulatory-acousticmodelsfor,Proctor2019ArticulatoryCO,Delattre1968ADS} which generates a lowered third formant. These configurations often co-occur with lip rounding, contributing to overall formant lowering. In contrast, non-rhotic British English dialects lack these tongue and lip postures specifically in word-final and pre-consonantal environments, leading to distinct articulatory patterns between the two dialects.  
\setlength{\parskip}{0pt}
Beyond rhoticity, dialectal differences exist in the back vowels, with British English generally exhibiting a more retracted tongue position compared to American English~\cite{ghorshi06_interspeech}. However, the extent to which tongue and lip posturing vary as a function of accent strengths remains underexplored, with acoustic-based analyses providing only indirect proxies for articulation.

\setlength{\parskip}{0pt}
Measuring speech articulation directly is resource-intensive and limited to laboratory environments with specialized instruments and generally small participant pools, making large-scale studies difficult. Electromagnetic articulography (EMA) provides direct kinematic measurements of articulators via flesh-point tracking, offering valuable insights into speech articulation and variability~\cite{wieling2016investigating,wieling2018analyzing}. However, its application in large-scale dialectal studies remains limited due to data scarcity and collection constraints. To address this, articulatory inversion models have been developed to estimate articulatory features from speech audio~\cite{elie24b_interspeech,fan24c_interspeech}. These models have been applied in articulation-grounded analysis of speech, as well as in downstream applications such as speech synthesis and accent adaptation. The availability of synchonous multi-modal corpora has  facilitated the development of acoustic-to-articulatory inversion methods, enabling the derivation of articulatory representations that enrich speech modeling by offering richer data grounded in the causal processes of speech production~\cite{cho2024codingspeechvocaltract,wu2023speaker,siriwardena2024accentconversionarticulatoryrepresentations}.  

In this work, we propose an automated approach to examining the relationship between accent strength and articulatory dynamics using articulatory inversion. Our framework jointly analyzes accent strength and articulatory features in a continuous manner, without reliance on discrete accent labels. To enhance interpretability, we reparameterize the predicted articulatory features from raw EMA coordinates into a less correlated and more linguistically meaningful feature space. Then, using phoneme-based distance measures as proxies for accent strength, we perform linear regression analyses to examine correlations between accent strength and these refined articulatory features. This approach allows for a more nuanced assessment of accent strength by capturing phonemic variation relative to specific pronunciation models. Our results reveal statistically significant correlations between certain articulatory features and accent strength, highlighting potential avenues for future research in accent modeling.

This study demonstrates how articulatory inversion enables large-scale articulatory analysis of accented speech. This method and its findings have implications for speech technology, accent adaptation, and language learning applications, providing insights that could inform pronunciation training tools, speech synthesis models, and computational accent modeling.

\section{Methods}

\subsection{Data}
We used a subset of the VCTK dataset~\cite{Veaux2017CSTRVC} for our experiment, comprising the three most represented accents in this data: 32 speakers from non-Scottish regions of the UK, referred to as British English and 21 American English speakers. 
Each speaker read approximately 400 utterances, contributing approximately 12 minutes of speech per speaker. The Montreal Forced Aligner (MFA)~\cite{mcauliffe2017montreal} was used to obtain word and phoneme alignments with the english\_us\_mfa and english\_uk\_mfa dictionaries and the english\_mfa acoustic model.

\subsection{Accent Strength Estimation}

Levenshtein distance (LD) between phonetic strings, also referred to as edit distance or phoneme error rate (PER), has been widely used to measure dialectal differences \cite{Kessler1995ComputationalDI}, with higher LD values indicating greater deviation from a reference pronunciation transcription. This study employs a weighted LD variant, following prior work that verifies a strong correlation of this measure with human ratings of accent strength~\cite{Wieling2014MeasuringFA} with performance on par with various acoustic measures~\cite{bartelds2021measuring}. For the weighted LD, a pointwise mutual information (PMI) score is computed for phoneme pairs as follows: 
\begin{align}\label{eq:PMI}
    PMI(x,y)&=\log_2\left(\frac{p(x,y)}{p(x)p(y)}\right)
\end{align} where $p(x,y)$ represents joint phoneme probabilities, and $p(x)$ and $p(y)$ are marginal probabilities within non-matching phoneme pair alignments. PMI scores are inverted, normalized, and iteratively refined until convergence to compute edit weights for weighted LD.

\begin{figure}
    \centering
    \includegraphics[width=0.9\linewidth]{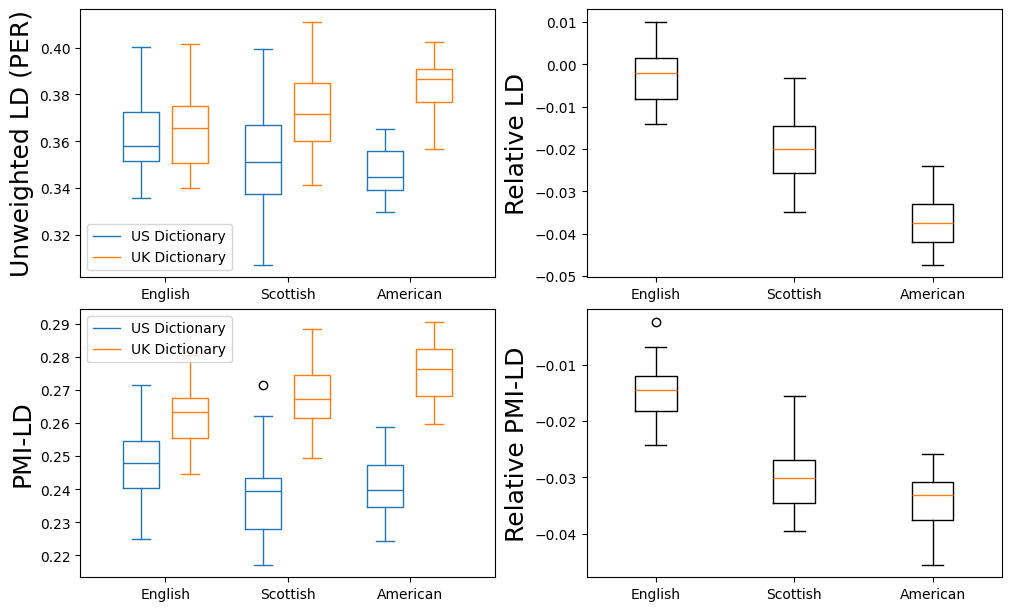}
    \caption{Unweighted LD (top) and PMI-LD (bottom) of phoneme prediction for each accent group against the US and UK dictionary (left), and the LD difference between the two dictionaries (right)}
    \label{fig:PER}
    \vspace{-1em}
\end{figure}

Although previous studies have used professionally transcribed phonetic annotations for PMI-LD, such annotations are costly and impractical for large-scale corpora. To address this, we use a Wav2Vec2-XLSR model fine-tuned on phoneme recognition~\cite{xu2021simpleeffectivezeroshotcrosslingual} as a noisy substitute for professional transcriptions. PMI-LD is computed as the distance between predicted phonemes and MFA phonemic transcriptions. LD and PMI-LD values are illustrated in Figure \ref{fig:PER}.

Not surprisingly, lower LD values are found on average for both the weighted and unweighted LD measures when the dictionary corresponding to the accent (i.e. dialect variation) of the speaker is used. However, deviations within accent categories are significantly reduced when considering \textit{relative} LD (calculated as the LD distance from UK dictionary subtracted from LD distance from US dictionary). 

\subsection{Articulatory Feature Estimation}

Articulatory features are estimated from acoustic speech recordings using the SPARC encoder~\cite{cho2024codingspeechvocaltract}, which predicts 12 EMA channels at 50Hz, simulating articulator fleshpoint tracking. These estimated channels provide normalized $x$, $y$ coordinates for the upper lip (UL), lower lip (LL), lower incisor (LI),  tongue tip (TT), tongue body (TB), and tongue dorsum (TD). 

\begin{figure}
    \centering
    \includegraphics[width=0.35\linewidth]{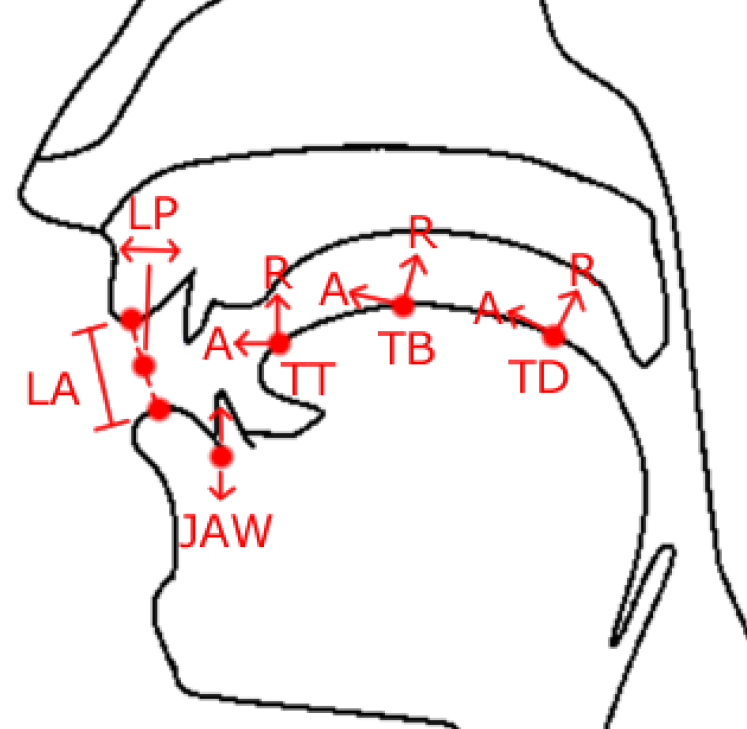}
    \caption{Reparameterized EMA channels}
    \label{fig:ema-points}
    \vspace{-1.5em}
\end{figure}

To extract linguistically relevant features, we compute lip aperture (LA) as the Euclidean distance between UL and LL, lip protrusion (LP) as the average $x$ value of UL and LL, and jaw height (JAW) as the $y$ value of LI. For the tongue sensors, we apply principal component analysis (PCA) to decorrelate the $x$, $y$ coordinates and define tongue advancement/retraction ($A$) and raising/lowering ($R$) as a new basis set $\{A,R\}$. As such, TTA denotes tongue tip advancement, TTR denotes tongue tip raising, and similar notation applies to other tongue sensors. Figure \ref{fig:ema-points} illustrates reparameterized EMA channels.

\begin{figure*}[t!]
    \centering
    \includegraphics[width=0.90\linewidth]{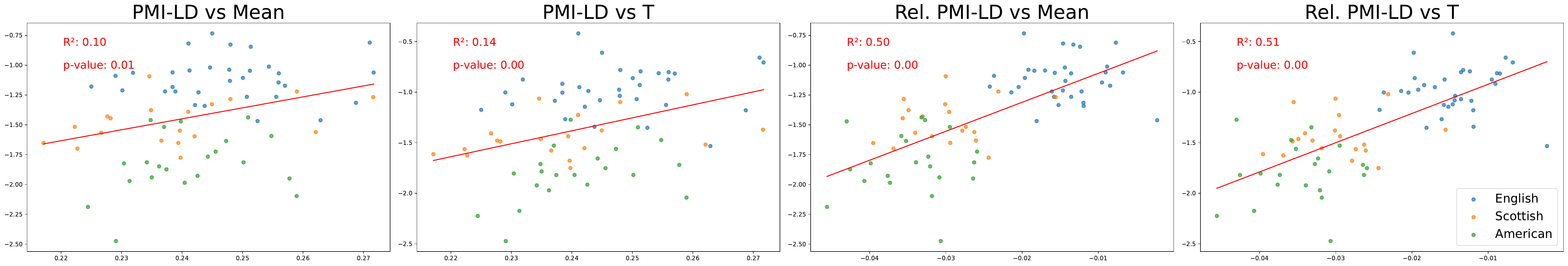}
    \caption{Linear regression plots for tongue tip advancement (TTA) for /\textipa{6}/. LD accent measure is plotted on the horizontal axis, and the mean articulatory feature is on the vertical axis. Linear regression lines and the corresponding $R^2$ and $p$-value are shown in red.}
    \label{fig:a-tongue-corrs}
    \vspace{-1em}
\end{figure*}
\subsection{Articulatory Task Dynamics Estimation}

To capture dynamic articulation, we estimate the equilibrium position value $T$ for each phoneme. The Task Dynamics model~\cite{saltzman1989dynamical} describes speech production as a sequence of linguistic gestures; each gesture is approximated by a damped harmonic oscillator~\cite{iskarous2017relation,simko2010embodied}. Under the simplifying assumption of a critically damped system with unit mass, the model is described by Equation \ref{eq:dynamic}, where $k$ represents stiffness and $T$ is the equilibrium position.
\begin{align}
    \frac{d^2x}{dt^2}+2\sqrt{k}\frac{dx}{dt}+k(x-T)=0\label{eq:dynamic}
\end{align}
For each phoneme, $T$ is estimated by optimizing the model to fit to the articulatory signal, minimizing the mean squared error (MSE) over a Gaussian-weighted rectangular window centered on the phoneme. The equation \ref{eq:dynamic} is analytically solved as:
\begin{equation}
\begin{aligned}
    r(t)&=\max(t-t_s,0)\\
    x(t)&=x_0 + (T-x_0)\left(1 - \left(1 + \sqrt{k}r(t)\right)e^{-\sqrt{k}r(t)}\right)
\end{aligned}
\label{eq:dyn-solved}
\end{equation}
where $x_0$ is the initial value (assuming zero derivative), and $t_s$ is the gesture onset. These parameters $T$, $k$, $x_0$, and $t_s$ are optimized using scipy's L-BFGS-B optimizer.


\subsection{Statistical Analysis}
\setlength{\parskip}{0pt}
To examine correlations between accent strength and articulatory features, we perform a simple linear regression between PMI-LD (US dictionary) and relative PMI-LD (US vs. UK dictionaries) as accent strength measures, and various statistical evaluations of the reparameterized EMA sensor trajectories. For simplicity, this paper focuses on tongue posturing and positioning associated with vowels and rhoticity, such as tongue advancement, and lip posturing and positioning associated with rounding, such as lip aperture and protrusion, as these are among salient interdialectal characteristics consistently reported in the literature ~\cite{hosseinzadeh2015british,hillenbrand1995acoustic,williams2014cross}. Using the acoustic phoneme alignment timings from MFA alignments, we extract the predicted EMA trajectories within each vowel segment boundary. Vowel segments are grouped by the vowel label used in the MFA transcript from the US dictionary to ensure vowels pronounced the same in one dictionary are grouped together, even if their actual phonetic realization differs across speakers. We consider correlation and regression results statistically significant at $p <0.05$.

\section{Results}

Our analysis reveals several significant correlations between specific articulatory features and accent strength. Figure \ref{fig:a-tongue-corrs} illustrates a sample of the linear regression analyses. Since PMI-LD values reflect distance from the US dictionary, positive correlations ($R>0$) indicate a stronger British accent, while negative correlations ($R<0$) correspond to a stronger American accent. Table \ref{tab:corr-summary} summarizes these significant correlations.

\subsection{Strongest Significant Correlations} 
Among the vowels examined, the strongest significant correlations are observed for the rhotic vowel /\textrhookrevepsilon/ and for the low back vowel /\textipa{6}/. For /\textrhookrevepsilon/, tongue dorsum raising (TDR) and tongue tip advancement (TTA) yield the highest absolute correlation values. Specifically, TDR (raising) shows a strong negative correlation with PMI-LD ($R=-0.85$ for mean values, $R=-0.84$, for $T$ (equilibrium values), while TTA (advancement) exhibits a strong positive correlation ($R=0.84$ for mean values, $R=0.81$ for $T$ equilibrium values). Additional correlations for /\textrhookrevepsilon/ include positive correlations for tongue advancement (TBA \& TDA), while negative correlations are found for tongue raising (TTR \& TBR), as well as $T$ for lip protrusion (LP), suggesting that British speakers produce /\textrhookrevepsilon/ with a more fronted tongue, while American speakers exhibit greater tongue raising.

For the low back vowel /\textipa{6}/, TTA (advancement) is positively correlated with relative PMI-LD ($R=0.70$ for mean values and $R=0.71$ for $T$ values). Moreover, additional positive correlations are observed for tongue body and dorsum advancement (TBA \& TDA), while a negative correlation for tongue body raising (TBR) is observed. This suggests that British speakers front the tongue more, whereas American speakers raise the tongue body more.

When using PMI-LD based solely on the US dictionary reference, significant correlations appear weaker ($|R|<0.5$) and are less frequent than those obtained with relative PMI-LD. On average, only 2.2 significant features per phoneme are found when using PMI-LD (US dictionary), compared to 5.4, when using relative PMI-LD (cross-dialectal). Given that PMI-LD has been established as a measure of accent strength, the remaining reported results focus on correlations with PMI-LD (US dictionary). 

\subsection{Additionally Significant Correlations Across Vowels}
Beyond /\textrhookrevepsilon/ and /\textipa{6}/, several other vowels exhibit articulatory patterns correlated with accent strength, although their correlations are generally weaker ($|R|<0.5$). For the high front tense vowel /\textipa{i}/, negative correlations are observed between PMI-LD for TDR and TBR mean values, while $T$ values are negatively correlated with TDR, TBA, and TDA. These findings suggest that for this vowel, American speakers may have a more raised tongue and fronted tongue body. The mid front vowel /\textipa{E}/ shows a weak positive correlation with the $T$ value of TTA, indicating a slight tendency toward a more fronted tongue tip in British speakers.


In the low and mid vowel categories, the front vowel /\ae/ shows a positive correlation with JAW and LP, suggesting greater jaw height and lip protrusion in American speakers, while a negative correlation with LA indicates that British speakers may produce /\ae/ with a smaller lip aperture. The high front lax vowel /\textipa{I}/ exhibits a positive correlation with the $T$ value of LA, indicating greater lip aperture in British speakers compared to American speakers.

For low back and rounded vowels, distinct patterns emerge. /\textipa{A}/ exhibits a weak positive correlation with mean TTR. But negative correlations with $T$ values of TBA, TDA, and TTA indicatie that American speakers may have a slightly more advanced tongue tip and tongue body for this vowel.

Among high and mid back vowels, /\textipa{0}/ and /\textipa{U}/ show negative correlations across multiple articulatory features. /\textipa{0}/ exhibits significant negative correlations with mean and $T$ values of TDR and TBR, indicating that American speakers may have a more raised tongue dorsum for this vowel. Similarly, /\textipa{U}/ shows negative correlations between PMI-LD and both mean and $T$ values of TBR, TDR, and JAW, suggesting that American speakers may produce this vowel with a more raised tongue dorsum and increased jaw height compared to British speakers. Additionally, /\textipa{U}/ exhibits negative correlations with the mean and $T$ values of LP, as well as the $T$ value of TDR, reinforcing differences in lip rounding and dorsum positioning between dialects.

\begin{table}[t]
\centering
\scriptsize
\setlength{\tabcolsep}{3pt}
\renewcommand{\arraystretch}{1}
\caption{Significant correlations between accent strength (PMI-LD), and mean ($\Bar{x}$) and equilibrium value ($T$) for each articulatory feature. Asterisks (*) indicate stronger correlations ($|R|>0.3$). Arrows indicate correlation sign (\textuparrow British, \textdownarrow American).}
\begin{tabular}{cc|ccc|ccc|c}
\toprule
            && TTA & TBA & TDA  & TTR & TBR & TDR & LA, LP, JAW \\
\midrule\midrule
/\textipa{i}/    &$\Bar{x}$ &-&-&-&-&$\downarrow$*&$\downarrow$*&-\\
                 &$T$       &-&$\downarrow$&$\downarrow$&-&$\downarrow$&$\downarrow$&-\\\cmidrule(lr){2-9}
/\textipa{6}/    &$\Bar{x}$ &$\uparrow$*&$\uparrow$&$\uparrow$&-&-&-&-\\
                 &$T$       &$\uparrow$*&$\uparrow$*&$\uparrow$*&-&$\downarrow$*&-&-\\\cmidrule(lr){2-9}
/\textipa{E}/    &$\Bar{x}$ &-&-&-&-&-&-&-\\
                 &$T$       &$\uparrow$&-&-&-&-&-&-\\\cmidrule(lr){2-9}
/\textipa{@}/    &$\Bar{x}$ &-&-&-&-&-&-&-\\
                 &$T$       &-&-&-&-&-&-&-\\\cmidrule(lr){2-9}
/\ae/            &$\Bar{x}$ &-&-&-&-&-&-&$\downarrow$LA* $\uparrow$LP* $\uparrow$JAW\\
                 &$T$       &-&-&-&-&-&-&-\\\cmidrule(lr){2-9}
/\textipa{I}/    &$\Bar{x}$ &-&-&-&-&-&-&-\\
                 &$T$       &-&-&-&-&-&-&$\uparrow$LA*\\\cmidrule(lr){2-9}
/\textipa{A}/    &$\Bar{x}$ &-&-&-&$\uparrow$&-&-&-\\
                 &$T$       &$\downarrow$&$\downarrow$&$\downarrow$&-&-&-&-\\\cmidrule(lr){2-9}
/\textipa{0}/    &$\Bar{x}$ &-&-&-&-&$\downarrow$*&$\downarrow$*&$\downarrow$JAW\\
                 &$T$       &-&-&-&-&$\downarrow$*&$\downarrow$*&$\downarrow$JAW\\\cmidrule(lr){2-9}
/\textipa{U}/    &$\Bar{x}$ &-&-&-&-&-&-&$\downarrow$LP\\
                 &$T$       &-&-&-&-&-&$\downarrow$*&$\downarrow$LP*\\\cmidrule(lr){2-9}
/\textrhookrevepsilon/  &$\Bar{x}$ &$\uparrow$*&$\uparrow$&$\uparrow$&$\downarrow$*&$\downarrow$*&$\downarrow$*&-\\
                        &$T$       &$\uparrow$*&$\uparrow$*&$\uparrow$*&$\downarrow$*&$\downarrow$*&$\downarrow$*&$\downarrow$LP\\
\bottomrule
\end{tabular}
\label{tab:corr-summary}
\vspace{-1.5em}
\end{table}

\subsection{Summary}
The results indicate that articulatory differences between American and British English speakers are most pronounced in the rhotic vowel /\textrhookrevepsilon/ and the low back vowel /\textipa{6}/, where tongue advancement and raising show the strongest correlations with accent strength. For /\textrhookrevepsilon/, British speakers exhibit a more fronted tongue tip and body, while American speakers show greater tongue rear (body \& dorsum) raising consistent with the articulatory properties of rhoticity. In /\textipa{6}/, British speakers advance the tongue tip and body more, whereas American speakers raise the tongue body more. Additional but weaker correlations are observed for most other vowels, suggesting that accent strength is less dramatically linked to phoneme-specific articulation in these cases.


\section{Discussion}

The results confirm that specific articulatory differences correlate with accent strength in British versus American English, particularly for /\textrhookrevepsilon/ and /\textipa{6}/, aligning with previous findings on rhoticity and back vowel in these English dialects. While correlations for other vowels are weaker, they suggest that variation throughout the vowel system is an underlying aspect of accent variation.

\subsection{Key Correlations in Articulatory Variation and Accent Strength}

The strongest correlations appear in the rhotic mid-central vowel /\textrhookrevepsilon/, where British speakers exhibit a more advanced tongue tip and tongue body, while American speakers show greater tongue dorsum raising. This pattern aligns with the non-rhotic nature of British English, in which British English typically lack the tongue retraction seen in American English rhotic productions~\cite{hosseinzadeh2015british,mielke2016individual,Proctor2019ArticulatoryCO}.

For the low back vowel /\textipa{6}/, the tongue tip and tongue body are more fronted in British speakers compared to American speakers. While this initially appears to contrast with previous formant-based studies, which associate British English with a lower F2 for back vowels, typically linked to a more retracted tongue position~\cite{ghorshi06_interspeech}, this discrepancy may reflect dialectal variation within British English, countervailing lip or pharynx posturing, or differences in measurement approaches. Prior studies have documented variability in back vowel articulation across British English dialects, which could contribute to these observed differences. 


For back rounded vowels (/\textipa{U}/, /\textipa{0}/), American speakers exhibit greater tongue dorsum raising and more lip protrusion, which is characteristic of a more rounded articulation. This aligns with previous findings that American English tends to maintain stronger rounding in high back vowels~\cite{williams2014cross}. However, for /\textipa{6}/ and /\textipa{U}/, our findings diverge from expectations based on F2-based studies, which typically associate British English with more lip rounding and tongue retraction for (at least certain) back vowels. These inconsistencies emphasize the need for complementary use of both articulatory and acoustic analyses, as well as consideration of prosodic and contextual effects in future research.

Because this study relies on articulatory inversion rather than direct articulatory measurements, some observed dialectal differences may stem from the lack of coarticulatory context analysis, the lack of articulatory information in EMA from the pharyx, larynx or velum, and/or potential limitations in the model’s ability to capture articulatory variability, particularly in dynamic speech settings.

\subsection{Evaluation of Accent Strength Measures and Articulatory Representations}

Relative PMI-LD yields more significant and higher correlations than PMI-LD, as Figure \ref{fig:PER} illustrates given the stronger separation between accent groups using the relative metric. However, while relative PMI-LD may better classify accents, its validity as a direct measure of accent strength remains to be fully investigated. Additionally, PMI-LD requires pronunciation dictionaries for both dialects, which may not be feasible for low-resource languages.

For articulatory analysis, both the mean and equilibrium target value ($T$) of each vowel segment show statistically significant correlations, with neither proving consistently superior. The mean is simpler and may be more robust to irregular trajectories, whereas $T$ offers a more interpretable estimate of an articulatory target. Future work should explore whether combining these measures enhances articulatory-accent correlations.

\subsection{Limitations and Future Directions}
A key limitation of this study is that vowel articulations were analyzed in isolation, without accounting for coarticulatory effects or prosodic context. Given that adjacent sounds influence articulatory shaping and positioning, future work should incorporate segmental and prosodic influences to better capture systematic variation. This is especially relevant for /\textipa{6}/, where unexpected tongue advancement in British speakers may stem from coarticulation rather than inherent articulatory differences.

Since this study relies on articulatory inversion rather than direct EMA measurements, validating these findings against real fleshpoint tracking data is essential. While the used articulatory inversion model can capture speaker-specific articulation with high correlation~\cite{cho2024codingspeechvocaltract}, differences between predicted and actual trajectories may introduce noise. Future studies should compare inversion-based predictions with direct EMA data to assess their reliability in accent research.

While this study identifies articulatory trends linked to accent strength, future research should explore how articulatory features interact with speech rhythm, lexical stress, reduction, and vowel phonological systems across dialects. Further refinement of accent strength measures, particularly relative PMI-LD, may improve the robustness of articulatory-accent correlations.



\section{Acknowledgements}
This work was supported by funds from USC Hearing, Communication and Neuroscience (HCN) pre-doctoral fellowship, NSF (IIS-2311676, BCS-2240349), and IARPA ARTS (award number 140D0424C0067, JHU subcontract) from the Office of the Director of National Intelligence.


\bibliographystyle{IEEEtran}
\bibliography{mybib}

\end{document}